\documentclass[12pt,letterpaper]{article}
\usepackage[margin=1in]{geometry}
\usepackage{amsmath,amssymb}
\usepackage{graphicx}
\usepackage{caption}
\usepackage{authblk}
\usepackage{float}
\usepackage{comment}
\usepackage{subcaption}
\usepackage{svg}
\usepackage{tikz}
\usepackage{adjustbox}
\usepackage{soul}
\usepackage{xcolor}
\usepackage{makecell}
\usepackage{booktabs}
\usepackage{multirow}
\usepackage{siunitx}
\usepackage{pdfpages}
\usepackage{array}
\usepackage[section]{placeins}

\let\oldsubsection\subsection
\renewcommand{\subsection}{\FloatBarrier\oldsubsection}
\let\oldsubsubsection\subsubsection
\renewcommand{\subsubsection}{\FloatBarrier\oldsubsubsection}

\usepackage[numbers,sort&compress]{natbib}
\usepackage{hyperref}

\title{Mechanistic Modeling of mRNA Precipitation in Batch and Continuous Operation via Population Balance Models (PBMs)}

\author[1]{Hasan Al-Mahayni}
\author[1]{Rajan~R.~Bhawnani}
\author[1]{Younghyeon Kim}
\author[1]{Krystian~Ganko}
\author[1]{Maria del Carme Pons Royo}
\author[1]{Allan S. Myerson}
\author[1,*]{Richard~D.~Braatz}

\affil[1]{Department of Chemical Engineering, Massachusetts~Institute~of~Technology, Cambridge, MA 02139, USA}
\affil[*]{Corresponding author: braatz@mit.edu}

\date{}

\begin{document}
\maketitle
\begin{center}
\section*{Abstract}
\end{center}
Messenger RNA (mRNA) therapeutics require scalable and cost-effective purification. Precipitation offers a promising alternative to chromatography, but its rational design requires understanding precipitate formation mechanisms. This article develops mechanistic population balance models (PBMs) for mRNA precipitation, calibrated using in-line concentration and chord length measurements for two constructs, COVID and Firefly Luciferase (FLuc). Conventional nucleation-and-growth models required unusually high supersaturation exponents to fit batch data and failed to predict the observed particle size response to reduced feed concentration. Instead, the data support aggregation-driven precipitate formation. A model combining primary nucleation, shear aggregation, and size-dependent breakage provides the best fit for both constructs according to the Akaike information criterion, with well-constrained posteriors for the estimated parameters. Size-dependent breakage improves model performance only together with shear aggregation, and their balance establishes a stable precipitate size. At half the feed concentration, the selected model reduces prediction error three-fold relative to Brownian aggregation and ten-fold relative to the best growth model. The five-fold larger COVID precipitates are associated with a 24-fold higher effective aggregation-to-breakage rate ratio. Without refitting, the batch-calibrated model predicts the outlet of a continuous tubular precipitator, with the outlet concentration within the tight prediction band and the mean chord length within 20\% of the measured value. These findings establish aggregation and breakage as key mechanisms governing mRNA precipitate size and demonstrate the predictive potential of batch-calibrated PBMs for continuous precipitation.
\bigskip

\noindent\textbf{Keywords:} mRNA precipitation; mRNA purification; population balance model; aggregation and breakage; continuous manufacturing; parameter estimation; uncertainty quantification

\section{Introduction}

Since the COVID-19 pandemic, mRNA-based therapeutics have rapidly emerged as a promising modality across a wide range of applications, including vaccines \cite{lu_safety_2024, banoun_mrna_2023}, gene therapy \cite{weissman_mrna_2015, lei_current_2020}, cancer immunotherapy \cite{qin_mrna-based_2022, miao_mrna_2021}, and regenerative medicine \cite{inagaki_cell_2024, wisitrasameewong_impact_2022}. As a result, global demand for mRNA production has increased substantially \cite{webb_current_2022}, creating an urgent need for scalable and cost-efficient continuous manufacturing platforms \cite{hengelbrock_scalable_2023, rosa_mrna_2021}. Downstream processing of mRNA often includes multiple affinity chromatography steps followed by tangential flow filtration (TFF) \cite{cui_comprehensive_2023} steps before lipid nanoparticle (LNP) encapsulation and freeze-drying \cite{schmidt_process_2022}. The chromatography steps are enabled by using resins such as oligo-dT \cite{cui_comprehensive_2023}. The challenge comes from the low binding capacity of oligo-dT (about 5 mg/mL), especially when compared to protein A chromatography for monoclonal antibodies, which has a binding capacity of 70 mg/mL \cite{grinsted_structure-performance_2025}. Low binding capacity and mass transfer limitations represent two important limitations of affinity chromatography, especially at high feed concentrations. Furthermore, other chromatography processes, such as size exclusion chromatography (SEC) and ion-pair reverse-phase chromatography (IPC), suffer from weak double-stranded RNA (dsRNA) removal or costly scale-up due to toxic reagents \cite{leong_revolutionizing_2025}. While continuous bioprocessing of active pharmaceutical ingredients (APIs) has been shown to reduce manufacturing costs, increase productivity, and decrease equipment size and cycle times \cite{burcham_continuous_2018, konstantinov_white_2015}, continuous column-based chromatography methods require synchronization of multiple columns to operate efficiently \cite{khanal_mathematical_2022}.

Continuous precipitation followed by TFF has emerged as a promising alternative to chromatographic purification and has been studied extensively for monoclonal antibodies (mAbs) \cite{yoshimoto_high-throughput_2013, li_continuous_2019, oelmeier_alternative_2013, li_enhanced_2021, hammerschmidt_economics_2014, burgstaller_continuous_2019, recanati_integration_2023, pons_royo_continuous_2023}. The main advantages of precipitation systems include simplicity, cheap reagents, scalability, and cost-effectiveness \cite{martinez_precipitation_2019}. Recently, Pons Royo et al.\ \cite{pons_royo_purification_2025} systematically evaluated precipitant composition and operating conditions for the precipitation of COVID and FLuc mRNA, demonstrating that polyethylene glycol (PEG) 6000 and sodium chloride (NaCl) at room temperature yield high purity and recovery in batch operation. Results showed yields of 81 to 85\% with purities of 80 to 82\% for FLuc and COVID mRNA, respectively. The precipitation performance was comparable to oligo-dT chromatography, although the purity was 10\% lower for precipitation. Building on these results, the subsequent study further showed that comparable performance can be achieved in a continuous dispersive plug-flow reactor (DPFR) equipped with static mixers and integrated with TFF \cite{pons_royo_continuous_2025}. Modelling the precipitation process in batch and continuous operation is crucial for process design, optimization, and control. In fact, to reduce the operational complexity of mRNA manufacturing, recent efforts have explored advanced control \cite{vetter_toward_2022, schmidt_process_2022}, automation \cite{schmidt_fast_2021}, and the development of digital twins \cite{schmidt_digital_2021, shahab_end--end_2025, helgers_digital_2021} for multiple unit operations. However, to the best of the authors' knowledge, no mechanistic PBM has been developed and experimentally validated for the precipitation of mRNA itself for purification purposes, in either batch or continuous operation.

Population balance models (PBMs) are widely used to describe crystallization and precipitation phenomena in bioprocesses \cite{ramkrishna_population_2014}. PBMs have been applied to study crystal polymorphisms \cite{flood_population_2013}, crystal morphology \cite{singh_comprehensive_2013}, tablet manufacturing \cite{zhang_development_2012}, and quality-by-design (QbD) strategies in pharmaceutical processes \cite{kayrak-talay_quality_2013}. Recently, the first PBM describing the nano-precipitation of mRNA into lipid nanoparticles (LNPs) was published \cite{shin_mechanistic_2025}.

A validated PBM coupled with an appropriate solubility model provides mechanistic insight into mRNA precipitation, enabling identification of the dominant particle formation mechanisms \cite{pessina_integrated_2025}. Quantifying model uncertainty is essential for assessing prediction confidence across operating conditions, including variations in mRNA concentration and precipitant composition \cite{streif_robustness_2016}. Such a framework also enables systematic comparison of precipitation behavior across mRNA constructs of different lengths and compositions, such as COVID and FLuc.

The primary objective of this work was to identify the dominant mechanisms governing mRNA precipitate formation and particle size evolution through the development and experimental validation of mechanistic PBMs. A second objective was to assess whether models calibrated using batch experiments could predict precipitation behavior in continuous operation without parameter refitting. If such transferability can be demonstrated, state estimators \cite{motz_state_2008,chen_linear_1984} and model predictive controllers (MPC) \cite{szilagyi_experimental_2019,su_integrated_2017,hermanto_integrated_2011} could be designed using only batch kinetic studies. The advantage of this approach is that batch experiments are less costly than continuous experiments, which require much larger processing volumes. Moreover, accurate prediction of the PSD at the DPFR outlet is valuable for downstream processing, as precipitate size strongly influences the performance of subsequent tangential flow filtration (TFF) steps \cite{minervini_optimizing_2024}.

In this study, we collect in-line concentration and chord length distribution (CLD) data in batch and compare two classes of PBMs for two mRNA constructs, COVID and FLuc. The first class describes particle formation by nucleation and growth, whereas the second describes formation by nucleation, aggregation, and breakage. We first select the fitting strategy and then compare the models with the Akaike information criterion (AIC). The growth models fit the data only with effective, unusually high supersaturation exponents. A model with shear aggregation and size-dependent breakage fit both constructs best, and the two mechanisms were only effective together. This result is consistent with previous work on protein precipitation. Precipitates of an Fc-fusion protein formed by aggregation of primary particles, and the large precipitates formed by further aggregation promoted by mixing \cite{greene_2018}. In population balance models of protein precipitation, the precipitate size is set by aggregation and shear-induced breakage \cite{raphael_rohani_1996,ehtesham_2025}. We quantifiy the parameter uncertainty of the selected model by Markov chain Monte Carlo (MCMC), compare the two constructs through their rates, and test the model on two conditions not used in fitting: a COVID batch at half the feed concentration and the outlet of a continuous tubular precipitator for FLuc.

\section{Theory}

\subsection{Population Balance and Mass Balances} \label{sec:Equations PBM}

The number density function $n(t,x)$ in a batch of constant volume $V$ evolves by nucleation, growth, aggregation, and breakage:
\begin{equation}
    \frac{\partial n}{\partial t} +\frac{\partial(Gn)}{\partial x} = B_0\,\phi(x) + \mathcal{A}[n] + \mathcal{F}[n],
    \label{eq:Batch dn/dt full}
\end{equation}
where $n$ is the particle length number density ($\mathrm{1/m^3/m}$), $x$ is the particle size (m), $t$ is the time (s), $B_0$ is the total nucleation rate ($1/\mathrm{m^3/s}$), $\phi(x)$ is the size distribution of the nuclei (1/m, $\int_0^\infty \phi\,\mathrm{d}x = 1$), $G$ is the growth rate ($\mathrm{m/s}$), and $\mathcal{A}[n]$ and $\mathcal{F}[n]$ are the aggregation and breakage terms. The kinetic expressions are given in Section~\ref{sec:theory aggregation}.

Aggregation and breakage are written in the particle volume $v = k_\mathrm{v}x^3$, with $n(v)\,\mathrm{d}v = n(x)\,\mathrm{d}x$ \cite{smoluchowski_1917}:
\begin{equation}
    \mathcal{A}[n](v) = \frac{1}{2}\int_0^{v} \beta(v-v',v')\,n(v-v')\,n(v')\,\mathrm{d}v' - n(v)\int_0^\infty \beta(v,v')\,n(v')\,\mathrm{d}v',
    \label{eq:aggregation}
\end{equation}
\begin{equation}
    \mathcal{F}[n](v) = \int_v^\infty S(v')\,b(v\,|\,v')\,n(v')\,\mathrm{d}v' - S(v)\,n(v),
    \label{eq:breakage}
\end{equation}
where $\beta$ is the aggregation kernel (m$^3$/s), $S$ is the breakage rate (1/s), and $b(v\,|\,v')$ is the daughter distribution, the number density of fragments of volume $v$ formed by the breakage of a particle of volume $v'$. In each equation, the first term is the birth of particles of volume $v$ and the second is their loss.

When nuclei are born at a single critical size, $\phi(x) = \delta(x-x_\mathrm{c})$, and \eqref{eq:Batch dn/dt full} without aggregation and breakage is equivalent to
\begin{equation}
    \frac{\partial n}{\partial t} + \frac{\partial(Gn)}{\partial x} = 0,
    \label{eq:Batch dn/dt reduced}
\end{equation}
with a nucleation boundary condition that treats the nuclei as a point source \cite{zhang_solving_2024}:
\begin{equation}
\left. nG \right|_{x = x_\mathrm{c}} = B_0.
\label{eq: BC xc}
\end{equation}
In all cases, there is no flux at infinite size,
\begin{equation}
\left. nG \right|_{x \to \infty} = 0,
\label{eq: BC_x inf}
\end{equation}
and the initial condition is $n(0,x)=0$ (no seed). The critical size was set to $x_{\mathrm{c}}=1$~\textmu{}m, the lower detection limit of the instrument used for measuring particle size (Section~\ref{sec: methods BSTR}).

Aggregation and breakage conserve the precipitated mass. The solute concentration $c(t)$ (g/L) is therefore consumed by nucleation and growth only:
\begin{equation}
    \frac{\mathrm{d} (cV)}{\mathrm{d} t}
    = - \rho k_{\mathrm{v}} V \!\left( B_{\mathrm{0}} \!\int_0^\infty \phi(x)\,x^3\,\mathrm{d}x + 3\! \int_{0}^{\infty} G n x^2 \mathrm{d}x \right),
    \label{eq:Batch dc/dt}
\end{equation}
where $\rho$ is the precipitate density and $k_{\mathrm{v}}$ is the volumetric shape factor \cite{zhang_solving_2024}. For $\phi = \delta(x-x_\mathrm{c})$, the first term reduces to $B_0 x_\mathrm{c}^3$. The precipitates were approximated as spheres ($k_\mathrm{v}=\pi/6$) to relate characteristic particle length to volume. The initial concentration $c(0)$ was set to the measured initial concentration $c_\mathrm{m}(t_i)$ (Section~\ref{sec: methods BSTR}). The precipitates sank in water, so their density was higher than that of water. Protein precipitates have densities of 1050 to 1300~$\mathrm{kg/m^3}$ \cite{bell_density_1982}, and $\rho = 1200$~$\mathrm{kg/m^3}$ was used.

For the DPFR, the axial position $z$ is added to the number density $n(t,x,z)$ and the solute concentration $c(t,z)$:
\begin{equation}
    \frac{\partial n}{\partial t}= -v_\mathrm{ax} \frac{\partial n}{\partial z} + D_\mathrm{ax} \frac{\partial^2 n}{\partial z^2} - \frac{\partial(Gn)}{\partial x} + B_0\,\phi(x) + \mathcal{A}[n] + \mathcal{F}[n],
    \label{eq:Continuous dn/dt}
\end{equation}
\begin{equation}
    \frac{\partial c}{\partial t}=-v_\mathrm{ax} \frac{\partial c}{\partial z}+ D_\mathrm{ax} \frac{\partial ^2 c}{\partial z^2} - \rho k_\mathrm{v} \biggl( B_0 \!\int_0^\infty \phi\, x^3\,\mathrm{d}x + 3 \!\int_0^\infty G n x^2\, \mathrm{d}x \biggr),
    \label{eq:Continuous dc/dt}
\end{equation}
where $v_\mathrm{ax}$ is the average axial velocity (m/s) and $D_\mathrm{ax}$ is the axial dispersion coefficient ($\mathrm{m^2/s}$). The boundary conditions in $x$ are the same as in batch. Danckwerts boundary conditions were applied to $n$ and $c$ in $z$. The feed contained no particles and had the solute concentration $c_\mathrm{feed}$:
\begin{equation}
\left[ n v_\mathrm{ax} - D_\mathrm{ax} \frac{\partial n}{\partial z} \right|_{z=0} = 0, \qquad
\left[ c v_\mathrm{ax} - D_\mathrm{ax} \frac{\partial c}{\partial z} \right|_{z=0} = v_\mathrm{ax}\,c_\mathrm{feed},
\end{equation}
\begin{equation}
\left. \frac{\partial n}{\partial z} \right|_{z=L_\mathrm{R}} = 0, \qquad
\left. \frac{\partial c}{\partial z} \right|_{z=L_\mathrm{R}} = 0.
\end{equation}
The axial velocity $v_\mathrm{ax}$ was the flow rate (0.5~mL/min) divided by the open cross-sectional area, that is, the cross-sectional area of the tube (8~mm$^2$) times the porosity. The porosity was set to 0.6, because the static mixers take up 40\% of the volume. This gave a mean residence time of $\tau = L_\mathrm{R}/v_\mathrm{ax} = 48$~min.

At steady state ($\partial/\partial t = 0$) and with negligible axial dispersion ($D_\mathrm{ax} = 0$), the DPFR reduces to a plug-flow reactor. With the residence time $\tau = z/v_\mathrm{ax}$, \eqref{eq:Continuous dn/dt} and \eqref{eq:Continuous dc/dt} become
\begin{equation}
    \frac{\partial n}{\partial \tau} + \frac{\partial(Gn)}{\partial x} = B_0\,\phi(x) + \mathcal{A}[n] + \mathcal{F}[n],
    \qquad
    \frac{\mathrm{d}c}{\mathrm{d}\tau} = - \rho k_\mathrm{v} \biggl( B_0 \!\int_0^\infty \phi\, x^3\,\mathrm{d}x + 3 \!\int_0^\infty G n x^2\, \mathrm{d}x \biggr),
    \label{eq:plug flow}
\end{equation}
with $n = 0$ and $c = c_\mathrm{feed}$ at $\tau = 0$. These are the batch equations \eqref{eq:Batch dn/dt full} and \eqref{eq:Batch dc/dt}, with time replaced by residence time. Both assumptions held for the continuous experiment in this work. Outlet samples were collected after approximately one nominal residence time. The high Péclet number ($\mathrm{Pe}\approx500$) supported the assumption of negligible axial dispersion and the use of a steady-state plug-flow approximation (see SI Section~S4.3).

\subsection{Kinetic Expressions and Model Sets}\label{sec:theory aggregation}

\subsubsection{Nucleation and Growth}\label{sec:kin nucleation}

Nucleation and growth were described by power laws in the relative supersaturation \cite{zhang_solving_2024,myerson_handbook_2002,randolph_chapter_1988,raphael_sunflower_1999}:
\begin{equation}
B_0=B_\mathrm{p}+B_\mathrm{s},
    \label{eq:B0}
\end{equation}
\begin{equation}
    B_\mathrm{p}=k_\mathrm{p} s^u ,
    \label{eq:Bp}
\end{equation}
\begin{equation}
    B_\mathrm{s} = k_\mathrm{b} s^b M,
\label{eq:Bs}
\end{equation}
\begin{equation}
    M = k_\mathrm{v} \rho \!\int_0^\infty n x^3  \mathrm{d}x,
\label{eq:M}
\end{equation}
\begin{equation}
    G = k_\mathrm{g} s^g (a+\gamma x)^p,
    \label{eq:G}
\end{equation}
\begin{equation}
    s = \frac{c-c_\mathrm{eq}}{c_\mathrm{eq}},
    \label{eq:S}
\end{equation}
where $c_\mathrm{eq}$ is the solubility of mRNA in the solvent (Section~\ref{sec: Materials and Methods}),  $B_\mathrm{p}$ is the primary nucleation rate with prefactor $k_\mathrm{p}$ and exponent $u$, $B_\mathrm{s}$ is the secondary nucleation rate with prefactor $k_\mathrm{b}$ and exponent $b$,
$G$ is the growth rate with prefactor $k_\mathrm{g}$ and exponent $g$, and the constants $a$, $\gamma$, and $p$ set its size dependence. Nuclei are born either at the critical size, $\phi = \delta(x-x_\mathrm{c})$, or with a log-normal size distribution centered at $x_\mathrm{c}$,
\begin{equation}
    \phi(x) = \frac{1}{x\,\sigma_\mathrm{n}\sqrt{2\pi}} \exp\!\left(-\frac{(\ln x - \ln x_\mathrm{c})^2}{2\sigma_\mathrm{n}^2}\right),
    \label{eq:PBE agg}
\end{equation}
with $\sigma_\mathrm{n} = 1$.

\subsubsection{Aggregation and Breakage}\label{sec:kin aggregation}

Two aggregation kernels are compared, Brownian (perikinetic) and shear (orthokinetic) aggregation \cite{smoluchowski_1917}:
\begin{equation}
    \beta_\mathrm{B}(x,x') = k_\mathrm{agg}\,\frac{2k_\mathrm{B}T}{3\mu}\,\frac{(x+x')^2}{x\,x'},
    \qquad
    \beta_\mathrm{S}(x,x') = \frac{4}{3}\,G_\mathrm{eff}\left(\frac{x+x'}{2}\right)^{3},
    \label{eq:kernels}
\end{equation}
where $k_\mathrm{B}$ is the Boltzmann constant, $T = 298$~K, $\mu = 8.9\times10^{-4}$~Pa\,s, $k_\mathrm{agg}$ is a dimensionless enhancement factor of the Brownian kernel, and $G_\mathrm{eff}$ (1/s) is an effective shear rate. $G_\mathrm{eff}$ is treated as a lumped rate constant that includes the shear rate, the collision efficiency, and effects of the precipitate structure that the model does not resolve, not as the physical shear rate. Breakage is binary, with fragments distributed uniformly in volume, $b(v\,|\,v') = 2/v'$. Size-independent ($\alpha = 0$) and size-dependent ($\alpha > 0$) breakage rates are compared,
\begin{equation}
    S(x) = S_0\!\left(\frac{x}{x_\mathrm{ref}}\right)^{\!\!3\alpha},
    \label{eq:breakage rate}
\end{equation}
with $x_\mathrm{ref} = 100$~\textmu{}m, where $S_0$ is the breakage rate of a 100~\textmu{}m particle.

\subsubsection{Model Sets}\label{sec:model sets}

Two classes of models were compared. Each was a special case of \eqref{eq:Batch dn/dt full}. The growth models had no aggregation or breakage ($\mathcal{A} = \mathcal{F} = 0$), and nuclei were born at the critical size ($\phi = \delta(x-x_\mathrm{c})$). Particles formed by nucleation and grew by deposition of solute. The parameter set $\Theta_4 := \{k_\mathrm{g}, g, k_\mathrm{p}, u\}$ covers primary nucleation and size-independent growth ($a = 1$, $\gamma = 0$). $\Theta_5$ adds size-dependent growth parameters ($\gamma$, with $a = p = 1$, as in the Canning and Randolph model \cite{canning_aspects_1967}), and $\Theta_6$ adds secondary nucleation parameters ($k_\mathrm{b}$, $b$) to $\Theta_4$.

The aggregation--breakage models had no growth and no secondary nucleation ($G = 0$, $B_\mathrm{s} = 0$), and nuclei were born with the log-normal distribution \eqref{eq:PBE agg}. Particles formed by primary nucleation and changed size only by aggregation and breakage. Growth was left out on purpose, to test whether aggregation and breakage alone could explain the measured size evolution. The two kernels and the two breakage rates gave four models, denoted $\Theta_4^{(K,\cdot)}$, with $K = \mathrm{B}$ (Brownian) or S (shear) and the second entry equal to $0$ (size-independent) or $\alpha$ (size-dependent breakage): $\Theta_4^{(\mathrm{B},0)} := \{k_\mathrm{p}, u, k_\mathrm{agg}, S_0\}$, $\Theta_4^{(\mathrm{B},\alpha)} := \{k_\mathrm{p}, u, k_\mathrm{agg}, S_0, \alpha\}$, $\Theta_4^{(\mathrm{S},0)} := \{k_\mathrm{p}, u, G_\mathrm{eff}, S_0\}$, and $\Theta_4^{(\mathrm{S},\alpha)} := \{k_\mathrm{p}, u, S_0, \alpha\}$ with $G_\mathrm{eff}$ fixed for COVID, or $\{k_\mathrm{p}, u, S_0, \alpha, G_\mathrm{eff}\}$ for FLuc (Section~\ref{sec: aggregation models}). The subscript 4 indicates that these models, like $\Theta_4$, have no secondary nucleation.

Shear aggregation and size-dependent breakage act on different sizes. The shear kernel grows with $(x+x')^3$, so large precipitates collide far more often than small ones. Size-dependent breakage ($\alpha > 0$) breaks large precipitates faster. A precipitate stops growing on average when aggregation and breakage balance. Section~\ref{sec: aggregation models} shows that only this combination gave a stable precipitate size whose dependence on the feed concentration was close to the data.

\subsection{Numerical Solution}\label{sec:Simulation}

All models were solved with an in-house finite volume (FV) solver in MATLAB, using the method of lines and the stiff solver \texttt{ode15s}.  Growth was discretized in flux form with a second-order MUSCL scheme and a van Leer limiter for non-uniform grids, which kept the number density non-negative (SI Section~S2.3.2). Aggregation and breakage were discretized with the fixed-pivot technique \cite{kumar_ramkrishna_1996}, which conserves particle volume exactly and gives the correct change in particle number in every event (one particle fewer per aggregation and one more per binary breakage; SI Sections~S2.3.4 and S2.3.5). For the aggregation--breakage models, an analytic Jacobian was supplied to the solver, which reduced the simulation time 6- to 30-fold, to about 1--2~s per simulation (SI Section~S2.3.6). All models in Table~\ref{tab:model_selection} were compared with the same numerics and data.

\subsection{Parameter Estimation}\label{sec:parameter estimation}

The parameters $\theta$ of each model were estimated by minimizing a weighted multi-objective least-squares function \cite{rippin_parameter_1978}, which combined the solute concentration and the PSD:
 \begin{equation}
 \label{eq:objective}
 \mathcal{J}(\theta)
 = \frac{w}{N_\mathrm{c}}\sum_{k=1}^{N_\mathrm{c}} \left( \frac{c(t_k,\theta)\!-\!c_{\mathrm{m}}(t_k)}{\sigma_{\mathrm{c}}(t_k)} \right)^{\!\!2}
 + \frac{1-w}{ML}\sum_{j=1}^{M}\sum_{\ell=1}^{L}\left(\frac{\mathcal{Q}(t_j,x_\ell,\theta)\!-\!\mathcal{Q}_{\mathrm{m}}(t_j,x_\ell)}{\sigma_{\mathcal{Q}}(t_j,x_\ell)}\right)^{\!\!2},
 \end{equation}
where $c$ and $c_\mathrm{m}$ are the model and measured concentrations at the times $t_k$, and $\mathcal{Q}$ and $\mathcal{Q}_\mathrm{m}$ are the model and measured size distributions at the times $t_j$ and bin midpoints $x_\ell$. The standard deviations $\sigma_\mathrm{c}$ and $\sigma_{\mathcal{Q}}$ were computed from the replicate runs. The factors $1/N_\mathrm{c}$ and $1/(ML)$ prevented the more frequently sampled data type from dominating, and $w \in (0,1)$ weighted the concentration against the PSD. For the CDF method, $w = 0.2$ was close to the elbow of the Pareto front between the two terms for both constructs (Figure~S2; SI Section~S2.4.3), and this value was used for all model comparisons.

The size distribution $\mathcal{Q}$ was either the PDF $f$ or the CDF $F$ of the particle counts in each bin, $N(t_j,x_\ell)=\int_{x_{\ell-1/2}}^{x_{\ell+1/2}} n\,\mathrm{d}x$:
\begin{equation}
f(t_j,x_\ell)=\frac{N(t_j,x_\ell)(\Delta x_\ell)^{-1}}{\sum_{i=1}^{L} N(t_j,x_i)}, \qquad
F(t_j,x_\ell)=\frac{\sum_{i=1}^{\ell} N(t_j,x_i)}{\sum_{i=1}^{L} N(t_j,x_i)},
\label{eq:PDF CDF}
\end{equation}
where $\Delta x_\ell$ is the bin width. The measured distributions were computed in the same way from the counts of the CLD probe. The CDF is less sensitive to bin-to-bin noise than the PDF \cite{berg_data_2008}. Because the CDF is cumulative, residuals in neighbouring bins are correlated. This correlation, and that between consecutive measurement times, was accounted for through the effective sample size $N_\mathrm{eff}$ (Section~\ref{sec: model selection criteria}). A third method fitted four normalized moments of the distribution (SI Section~S2.4.1). The three methods were compared in Section~\ref{sec: Model fitting and method comparison}. 

$\mathcal{J}$ was minimized with the MATLAB \texttt{lsqnonlin} function within the bounds of Table~\ref{tab:parameter_bounds}, with all prefactors ($k_\mathrm{g}$, $k_\mathrm{p}$, $k_\mathrm{b}$, $k_\mathrm{agg}$, $G_\mathrm{eff}$, $S_0$) estimated as $\log_{10}$ values. Each model was fitted once from an initial guess and then from 100 random starts, and the best fit was kept. An example of convergence of the multi-start is shown in Figure S1 (SI Section~S2.4.2). Approximate 95\% confidence intervals were computed from the Jacobian at the optimum, with the residual variance scaled by an effective sample size $N_\mathrm{eff} = 100$ that accounted for the correlation of neighbouring residuals in time and size (SI Section~S2.7.1).

The bounds were set wider than the ranges reported for crystallization, because mRNA might follow neither small-molecule ($u \in [5,15]$, $b \in [1,4]$, $g \in [1,2]$ \cite{myerson_handbook_2002,jw_mullin_crystallization_2001}) nor protein kinetics ($u$ of 10--20 \cite{nagy_internal_2011,le_minh_simple_2022,ozaki_population_2019}). The upper bounds of $g$ and $b$ were consequently increased from 2 to 4 and from 4 to 5, respectively (Table~\ref{tab:parameter_bounds}).
\label{sec: Priors from literature}

\begin{table}[!htbp]
\centering
\caption{Parameter bounds used in parameter estimation (Section~\ref{sec:parameter estimation}) and as uniform priors in MCMC (Section~\ref{sec: MCMC theory}).}
\begin{tabular}{ccc}
\toprule
\textbf{Parameter} & $\theta_{i,\mathrm{min}}$ & $\theta_{i,\mathrm{max}}$ \\
\midrule
\multicolumn{3}{c}{Growth models $\Theta_4$--$\Theta_6$} \\
\midrule
$\log_{10}k_g$ & $-15$ & $-5$  \\
$g$   & $1$   &  $4$ \\
$\log_{10}k_p$ &  $-40$ & $5$ \\
$u$   & $5$ & $20$  \\
$\log_{10}k_b$ &  $-40$ & $12$  \\
$b$   & $1$ & $5$  \\
$\log_{10}\gamma$ & $-2$ & $8$ \\
\midrule
\multicolumn{3}{c}{Aggregation--breakage models $\Theta_4^{(K,\cdot)}$} \\
\midrule
$\log_{10}k_p$ & $-5$ & $10$ \\
$u$   & $1$ & $10$ \\
$\log_{10}k_\mathrm{agg}$ & $5$ & $14$ \\
$\log_{10}G_\mathrm{eff}$ & $1$ & $10$ \\
$\log_{10}S_0$ & $-4$ & $9$ \\
$\alpha$ & $0$ & $1.5$ \\
\bottomrule
\end{tabular}
\label{tab:parameter_bounds}
\end{table}

\subsection{Markov Chain Monte Carlo-based Uncertainty Analysis} \label{sec: MCMC theory}

Linearized confidence intervals can be misleading for nonlinear models \cite{berliner_nonlinear_2021}. The posterior of the selected model $\Theta_4^{(\mathrm{S},\alpha)}$ was therefore sampled by MCMC, with the affine-invariant ensemble sampler (stretch move) of \texttt{emcee} \cite{foreman-mackey_emcee_2013}, implemented in MATLAB. The priors were uniform within the bounds of Table~\ref{tab:parameter_bounds}, with $k_\mathrm{p}$, $S_0$, and $G_\mathrm{eff}$ sampled in $\log_{10}$ space. The log-likelihood was
\begin{equation}
 \ln L(\theta) = -\frac{N_\mathrm{eff}}{2}\,\ln \mathcal{J}(\theta, w, c_\mathrm{m}, t_k, \mathcal{Q}_\mathrm{m}, t_j, x_\ell),
 \label{eq:likelihood}
\end{equation}
with $\mathcal{J}$ from \eqref{eq:objective} at $w = 0.2$. Eq.~\eqref{eq:likelihood} treats $\mathcal{J}$ as the sum of squares of $N_\mathrm{eff}$ independent Gaussian residuals, with the unknown noise variance profiled out (SI Section~S2.5.1). It used the same $N_\mathrm{eff}$ as the confidence intervals and the AIC (Section~\ref{sec: model selection criteria}), so the three were consistent. Parameter sets outside the bounds, or giving a failed simulation, were assigned $\ln L = -\infty$.

One chain ensemble was run per construct: four parameters for COVID ($G_\mathrm{eff}$ fixed) and five for FLuc. Twenty-four walkers were started from small Gaussian perturbations of the least-squares estimate, scaled by the linearized covariance. Each ensemble ran for 4000 steps. The first 1200 steps were discarded as burn-in, which left 67{,}200 samples per construct. The 95\% credible intervals changed by less than 0.06 ($\log_{10}$ units) between 2000 and 4000 steps.

\subsection{Forward Uncertainty Propagation from Parameters to Outputs} \label{sec: theory output propogation}

Parameter uncertainty was propagated to the outputs by Monte Carlo sampling of the posterior. From the 67{,}200 posterior samples, 2000 were drawn at random and simulated with the full model, for both the batch and the continuous predictions. The resulting concentration and PSD trajectories were summarized by their median and their 68\% (16th--84th percentile) and 95\% (2.5th--97.5th percentile) bands.

\subsection{Model Selection Criteria}\label{sec: model selection criteria}

The growth models and the aggregation--breakage models are not nested. They were therefore compared with the Akaike information criterion (AIC) \cite{akaike_1974} instead of an F-test, computed from the likelihood of Eq.~\eqref{eq:likelihood} (SI Section~S2.5.1):
\begin{equation}
\mathrm{AIC} = N_\mathrm{eff}\,\ln\!\left(\frac{\mathcal{J}}{N_\mathrm{eff}}\right) + 2N_\mathrm{p},
\label{eq:AIC}
\end{equation}
where $N_\mathrm{p}$ is the number of fitted parameters and $N_\mathrm{eff} = 100$. The model with the lowest AIC was preferred. Because $\mathcal{J}$ is a weighted objective and not an exact likelihood, the AIC was used only to rank the models. The selected model was the same for any $N_\mathrm{eff} > 13.1$ (SI Section~S2.7.2). The candidate models were the growth models $\Theta_4$--$\Theta_6$ and the four aggregation--breakage models $\Theta_4^{(K,\cdot)}$ (Section~\ref{sec:theory aggregation}). In addition to the fit, the models were judged by two predictions not used in fitting: a COVID batch at a lower feed concentration (0.36~g/L) and the outlet of the continuous FLuc precipitator.

\section{Materials and Methods}\label{sec: Materials and Methods}
\subsection{mRNA Constructs and Reagents}\label{sec: mRNA constructs}
Experiments were performed using purified FLuc mRNA (2117 nt, 660 kDa, 1 g/L) (Arranta Bio, Recipharm AB, USA) and purified COVID mRNA (4283 nt, 1384 kDa, 1 g/L) (Arranta Bio, Recipharm AB, USA), with samples stored at $-20^\circ\mathrm{C}$ for short-term use and $-80^\circ\mathrm{C}$ for long-term preservation.

\subsection{Chemicals}\label{sec: Chemicals}
All reagents employed in this study were of analytical grade and obtained from Sigma-Aldrich (St.\ Louis, MO, USA) or MilliporeSigma (Burlington, USA), unless stated otherwise. Solutions were prepared using RNase-free water from the Direct-Q\textsuperscript{\textregistered} 3 Remote Kit with Biopak Final Filter or Nuclease- and Endotoxin-Free Water from MilliporeSigma (Burlington, USA).

\subsection{Batch Stirred Tank Reactor} \label{sec: methods BSTR}

The batch precipitation of mRNA was performed by mixing an mRNA solution (Arranta Bio, Recipharm AB, USA) with a precipitant (40\% w/w PEG6000, 0.8 M NaCl) in a 7:3 volumetric ratio, resulting in a total volume of 1.5 mL and initial conditions of approximately 0.7 g/L mRNA, 13\% (w/w) PEG6000 and 250 mM NaCl. The concentration of precipitated mRNA was measured every 3 seconds using a NanoDrop\textsuperscript{TM} One\textsuperscript{\textcopyright} Spectrophotometer (Thermo Scientific\textsuperscript{TM}, USA), based on a linear relationship between the concentration of precipitated solid mRNA and the absorbance at 500 nm, which represents the turbidity of the system \citep{Bhawnani2026}. The concentration of dissolved mRNA was then back-calculated by mass balance. In parallel, the Blaze 900 probe (BlazeMetrics\textsuperscript{TM}, USA) was used to obtain real-time images and chord length distribution (CLD) data during precipitation in a custom-built 3~mL batch reactor with a working volume of 1.5 mL. CLD data were sampled every 5 seconds. A schematic of the setup is shown in Figure~S3. Batch experiments were run in triplicate for COVID and in duplicate for FLuc. One additional COVID batch with a feed concentration of 0.36~g/L, at the same PEG6000 and NaCl concentrations, was used only for validation.

In addition to in-line measurements, the concentration of mRNA in the liquid phase before and after the precipitation was measured off-line using the NanoDrop\textsuperscript{TM} One\textsuperscript{\textcopyright} Spectrophotometer to determine the initial mRNA concentration $c_\mathrm{m}(t_i)$, the final concentration of mRNA remaining in the liquid phase after precipitation $c_\mathrm{m}(t_f)$, and the equilibrium concentration (or solubility) after 48 hours ($c_\mathrm{eq}$). The liquid-phase concentration of mRNA after precipitation $c_\mathrm{m}(t_f)$ or $c_\mathrm{eq}$ was measured from the supernatant after centrifugation for 12 minutes at 15,000 relative centrifugal force. The solubility $c_\mathrm{eq}$ of mRNA was previously determined under different concentrations of NaCl salt and PEG6000, including the previously described operating conditions \cite{pons_royo_purification_2025}. Since the results in the study showed $>$95\% yields for both mRNA constructs, $c_\mathrm{eq}$ was chosen to be 5 \textmu{}g/mL (or 0.005 g/L) in this work, considering that the lower detection limit of RNA concentration using the NanoDrop\textsuperscript{TM} One\textsuperscript{\textcopyright} Spectrophotometer is 1.6 \textmu{}g/mL \cite{thermo_fisher_scientific_inc_nanodrop_2021}. These numbers were consistent with the measured values of $c_\mathrm{m}(t_f)$ and $c_\mathrm{eq}$ in this study.

\subsection{Continuous Dispersive Plug Flow Reactor} \label{sec: DPFR methods}

Continuous precipitation of mRNA was performed using a custom tubular reactor setup. The reactor consisted of Tygon\textsuperscript{\textregistered} R-3603 tubing (3.20 mm inner diameter; Avantor, USA) fitted with Acetal static mixing elements (HT-40-3.18-12-AC; Stamixco AG, New York, USA) and connected through Luer-lock fittings (Cole-Parmer). The reactor had a total length $L_\mathrm{R}$ of five meters. The mRNA solution (Arranta Bio, Recipharm AB, USA) was fed at 0.35 mL/min and combined in line with a precipitant stream (40\% w/w PEG6000, 0.8 M NaCl) delivered at 0.15 mL/min, resulting in final in-reactor conditions of 13\% (w/w) PEG6000 and 250 mM NaCl \cite{pons_royo_continuous_2025}. Under these operating conditions, the reactor provided a residence time of approximately 48 minutes. The outlet suspension was collected 10~min after one residence time had elapsed (58~min after start-up). The collected outlet was further characterized with the Blaze 900 and the
NanoDrop\textsuperscript{TM}. The mRNA concentration in the liquid phase of the collected outlet was measured off-line and was 0.01~g/L. Due to the larger processing volumes required to run the continuous setup, only one experiment was performed in this case. Finally, due to material constraints, only FLuc was precipitated in the DPFR.

\subsection{Chord Length Distribution and Particle Size Distribution} \label{sec: CLD to PSD}

The CLD was measured with the Blaze 900 probe, which uses image-derived focused beam reflectance measurement (ID-FBRM) with logarithmically spaced bins. Converting a CLD to a PSD is an ill-posed inverse problem: small measurement noise and wrong shape or bin assumptions can produce large errors in the reconstructed PSD \cite{hukkanen_measurement_2003,li_determination_2005,kempkes_measurement_2008}. No validated inversion exists for the Blaze probe, and an accurate inversion would require the precipitate shape, which was irregular here. Togkalidou et al.\ \cite{togkalidou_parameter_2004} instead fitted kinetic parameters directly to the low-order moments of the CLD measured with a Lasentec\textsuperscript{\textregistered} FBRM probe. The resulting parameters were essentially the same as those estimated from the PSD, except for one nucleation prefactor, because the CLD and the PSD follow similar dynamic trends. CLD data at small sizes should be interpreted with care, because of the detection limit and surface-roughness artifacts \cite{gunawan_identification_2002, hukkanen_measurement_2003,togkalidou_parameter_2004}. In this study, the PBM was fitted directly to the CLD, which gave a gray-box model \cite{togkalidou_parameter_2004}. Gray-box models are widely used in process control, because they need limited data and extrapolate better than purely data-driven models \cite{simpkins_system_2012}.

\section{Results and Discussion}

\subsection{Batch Precipitation Dynamics} \label{sec: batch data}

\begin{figure}[!htbp]
\centering
\includegraphics[width=\textwidth]{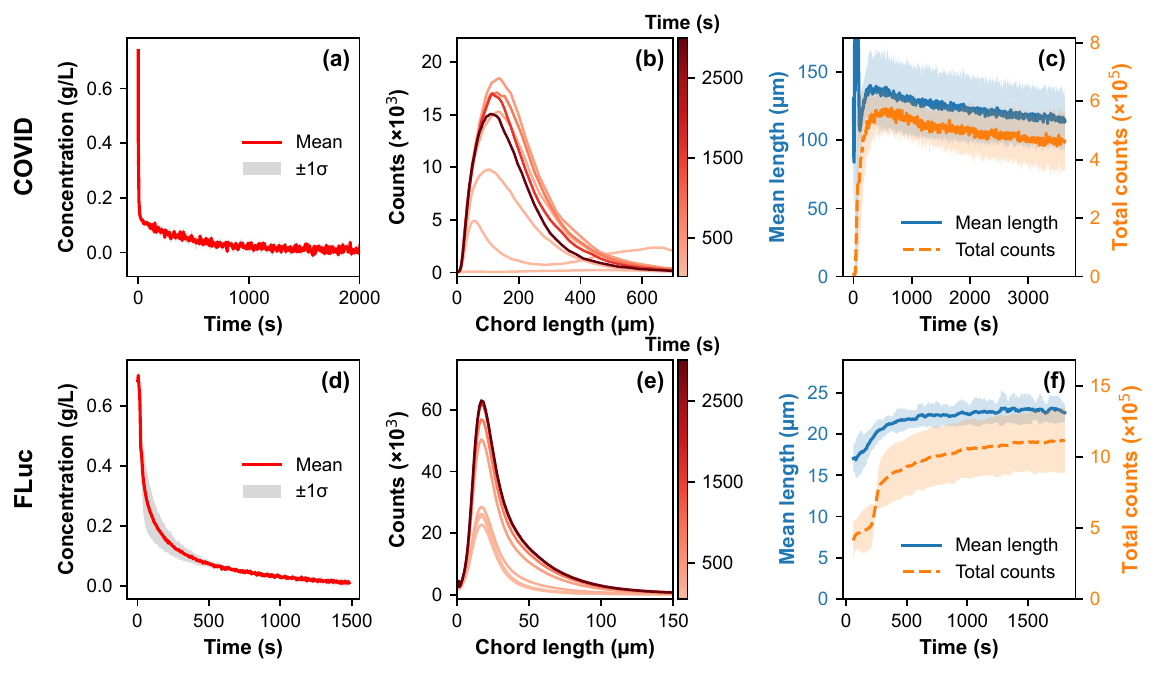}
\caption{Batch precipitation data for COVID (top row) and FLuc (bottom row). (a, d) Mean mRNA concentration in the liquid, with the $\pm1\sigma$ band across runs. (b, e) Mean CLD at selected times; darker lines are later times. (c, f) Mean chord length (left axis) and total counts (right axis) versus time, with $\pm1\sigma$ bands.}
\label{fig:data}
\end{figure}

Figure~\ref{fig:data} shows the batch precipitation data for COVID and FLuc mRNA. The liquid-phase concentration decreased shortly after the precipitant was added ($t = 0$~s; Figure~\ref{fig:data}ad), more slowly for FLuc than for COVID. The equilibrium concentration measured off-line was about 0.005~g/L, corresponding to more than 99\% conversion. The mean chord length and total counts (Figure~\ref{fig:data}cf) followed similar dynamics, and COVID reached steady values faster than FLuc. The COVID precipitates were about five times larger (mean chord length about 125~\textmu{}m, compared with 22~\textmu{}m for FLuc). For COVID, the mean chord length peaked at about 500~s and then declined slightly (Figure~\ref{fig:data}c), after the solute was nearly depleted. A size decrease at almost constant solid mass suggested breakage of the large COVID precipitates. The CLDs (Figure~\ref{fig:data}be) were approximately log-normal; their peaks rose and their right tails lengthened over time. The precipitates were amorphous, with no well-defined shape (Blaze images in SI Figure~S4). Early CLD measurements had a low signal-to-noise ratio, because the precipitates were few and small and acquisition was disturbed by mixing. FLuc also showed many counts near the 1~\textmu{}m detection limit, where measurements are unreliable \cite{xu_particle_2002} (Section~\ref{sec: CLD to PSD}). CLD data before 60~s were therefore excluded from the analysis.

\subsection{Model Fitting and Prediction on Batch Data} \label{sec: batch modelling}

\subsubsection{Fitting Method Comparison} \label{sec: Model fitting and method comparison}

\begin{figure}[!htbp]
\centering
\includegraphics[width=\textwidth]{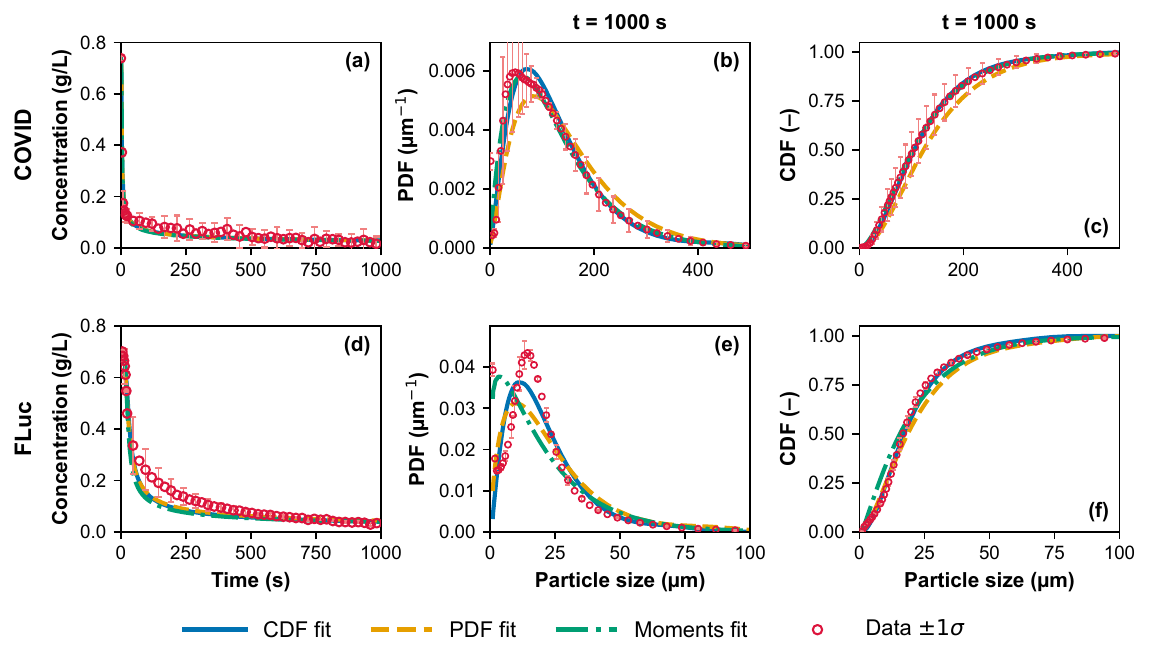}

\vspace{-0.5cm}

\caption{Fits of the growth model $\Theta_6$ with the three methods (CDF, PDF, and moments), for COVID (top row) and FLuc (bottom row): (a, d) concentration versus time, and (b, e) PDF and (c, f) CDF at $t = 1000$~s. Data are shown as mean $\pm1\sigma$.}
\label{fig:METHOD_COMPARISON}
\end{figure}

The three fitting methods were compared with the growth model $\Theta_6$. For the PDF and moment fits, $w$ was taken at the elbow of each Pareto front in Figure S2. For the CDF fits, $w = 0.2$ was used for both constructs, where both CDF fronts were nearly flat (Figure~S2). The fitted parameters are listed in Table~\ref{tab:method_comparison}. All three methods reproduced the concentration similarly (Figure~\ref{fig:METHOD_COMPARISON}ad) but differed in the distributions. For both constructs, PDF fitting gave a poorer distribution at $t = 1000$~s than the CDF and moment fits (Figure~\ref{fig:METHOD_COMPARISON}bc). For FLuc, moment fitting missed the peak and shape of the distribution (Figure~\ref{fig:METHOD_COMPARISON}ef), so matching a few moments did not ensure a good fit of the full distribution (Figure~S5). CDF fitting was the only method that worked well for both constructs and was used for all further analyses, with $w = 0.2$.

\begin{table}[!htbp]
\centering
\caption{Parameters of the growth model $\Theta_6$ fitted with the three methods. The bold rows correspond to the selected method (CDF). Values marked with $^\ast$ lie at a bound of Table~\ref{tab:parameter_bounds}.}
\begin{tabular}{lccccccc}
\toprule
\textbf{Method} & $w$ & $\log_{10}k_g$ & $g$ & $\log_{10}k_p$ & $u$ & $\log_{10}k_b$ & $b$ \\
\midrule
\multicolumn{8}{c}{\textbf{COVID}} \\
\midrule
\textbf{CDF}     & $\mathbf{0.2}$ & $\mathbf{-11.32}$ & $\mathbf{3.49}$ & $\mathbf{-17.30}$ & $\mathbf{11.10}$ & $\mathbf{-1.64}$ & $\mathbf{4.98}$ \\
PDF              & $0.5$  & $-11.28$ & $3.52$ & $-24.62$ & $14.20$ & $-1.84$ & $5.00^\ast$ \\
Moments          & $0.5$  & $-11.29$ & $3.54$ & $-5.17$  & $5.00^\ast$ & $-0.87$ & $4.61$ \\
\midrule
\multicolumn{8}{c}{\textbf{FLuc}} \\
\midrule
\textbf{CDF}     & $\mathbf{0.2}$ & $\mathbf{-11.87}$ & $\mathbf{2.95}$ & $\mathbf{-2.50}$ & $\mathbf{5.00^\ast}$ & $\mathbf{1.41}$ & $\mathbf{4.19}$ \\
PDF              & $0.7$  & $-12.00$ & $3.07$ & $-2.68$  & $5.00^\ast$ & $1.82$ & $3.85$ \\
Moments          & $0.7$  & $-11.90$ & $3.13$ & $-3.66$  & $5.00^\ast$ & $2.96$ & $3.34$ \\
\bottomrule
\end{tabular}
\label{tab:method_comparison}
\end{table}

\subsubsection{Growth Models}\label{sec: Model reduction}

\begin{figure}[!htbp]
\centering
\includegraphics[width=0.74\textwidth]{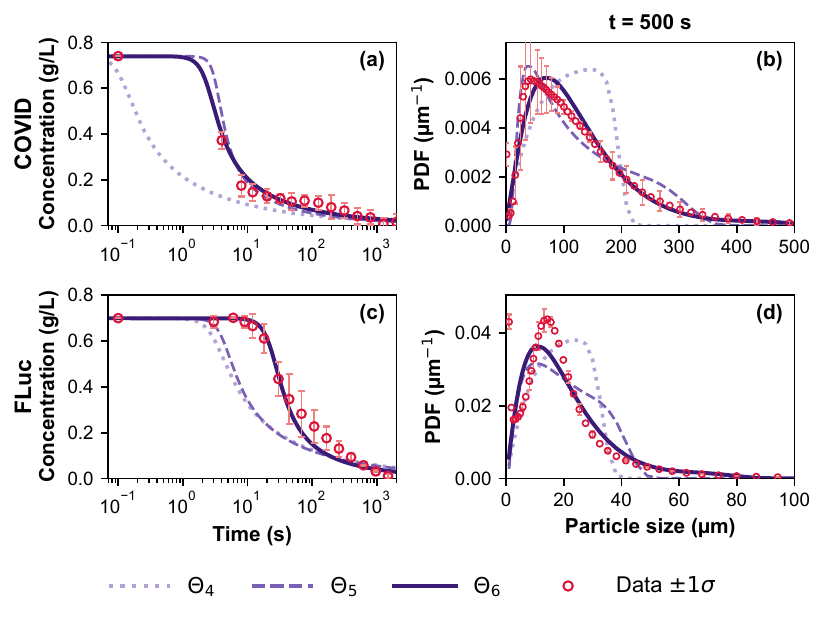}
\caption{Fits of the growth models $\Theta_4$, $\Theta_5$, and $\Theta_6$ for COVID (top row) and FLuc (bottom row): (a, c) concentration versus time (logarithmic time axis) and (b, d) PDF at $t = 500$~s. Data are shown as mean $\pm1\sigma$ across runs (three for COVID, two for FLuc); the data were down-sampled for clarity.}
\label{fig:model_reduction}
\end{figure}

Figure~\ref{fig:model_reduction} compares the growth models $\Theta_4$, $\Theta_5$, and $\Theta_6$. The AIC improved from $\Theta_4$ to $\Theta_6$ for both constructs (Table~\ref{tab:model_selection}): from $-455$ to $-521$ for COVID and from $-200$ to $-378$ for FLuc. Secondary nucleation was therefore the most useful addition to the growth model. $\Theta_4$ and $\Theta_5$ gave distorted PDFs at 500~s, with a sharp cut-off at large sizes or a shoulder (Figure~\ref{fig:model_reduction}bd). $\Theta_6$ reproduced the PDF shape for both constructs. $\Theta_4$ and $\Theta_5$ depleted the FLuc solute too fast between about 3 and 200~s (Figure~\ref{fig:model_reduction}c). $\Theta_6$ followed the FLuc concentration more closely, but the shape of its curve differed from the data. It fell more steeply between about 20 and 50~s and then levelled off, whereas the measured concentration kept declining until the end of the batch. This late decline sets the final conversion, so it matters most when the model is extrapolated to longer times, as in the continuous precipitator (Section~\ref{sec:Cont}).

The best growth model $\Theta_6$ needed growth and secondary nucleation exponents of $g \approx 2.9$--$3.5$ and $b \approx 4.4$--$5$, above the usual ranges (Section~\ref{sec: Priors from literature}). Such exponents are effective values, as they suggested that the model compensated for a mechanism that it did not include. The COVID data pointed to this mechanism: the mean chord length decreased after 500~s at almost constant solid mass (Figure~\ref{fig:data}c), which a growth model cannot reproduce. Aggregation and breakage were therefore tested in Section~\ref{sec: aggregation models}.

\subsubsection{Aggregation--Breakage Models}\label{sec: aggregation models}

\begin{figure}[!htbp]
\centering
\includegraphics[width=\textwidth]{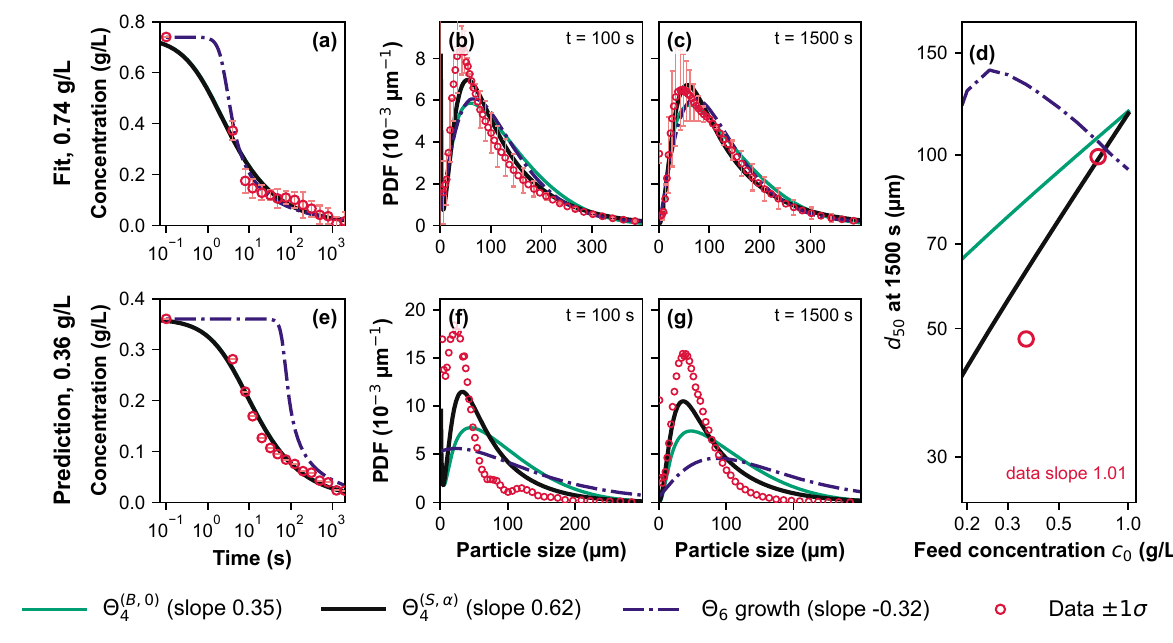}
\caption{Aggregation--breakage models for COVID, fitted at 0.74~g/L (top row) and predicted at 0.36~g/L (bottom row; not used in fitting): (a, e) concentration versus time (logarithmic time axis), (b, f) PDF at $t = 100$~s, and (c, g) PDF at $t = 1500$~s. (d) Number median size $d_{50}$ at 1500~s versus the feed concentration $c_0$ (log--log), with the slope of each model in the legend; circles are the two measured points. The best growth model $\Theta_6$ (dash-dot), fitted and predicted in the same way, is shown for reference. Data are shown as mean $\pm1\sigma$ (one run at 0.36~g/L, no error bars).}
\label{fig:aggregation models}
\end{figure}

For COVID, the four aggregation--breakage models of Section~\ref{sec:theory aggregation} were fitted at 0.74~g/L and then used to predict the batch at 0.36~g/L without refitting (Figure~\ref{fig:aggregation models}). Table~\ref{tab:model_selection} summarizes the results. All aggregation--breakage models that could be fitted improved on $\Theta_4$ and $\Theta_5$, and the Brownian models reached about the same AIC as the best growth model $\Theta_6$, with fewer parameters. At 0.74~g/L, all fitted models described the PDF at 100 and 1500~s about equally well, including $\Theta_6$ (Figure~\ref{fig:aggregation models}bc). The fit of the size distribution alone therefore could not discriminate between growth and aggregation. The concentration plots, however, showed that the aggregation--breakage models followed the measured depletion, whereas $\Theta_6$ stayed near the feed concentration for the first seconds and then fell below the data (Figure~\ref{fig:aggregation models}a). The prediction at 0.36~g/L discriminated more clearly, as discussed below. The combination of shear aggregation and size-dependent breakage ($\Theta_4^{(\mathrm{S},\alpha)}$) gave the best fit. $\Theta_4^{(\mathrm{S},\alpha)}$ reached $\mathcal{J} = 0.447$ and AIC $= -533.1$, compared with $-520.3$ for $\Theta_4^{(\mathrm{B},0)}$ and $-520.9$ for the best growth model $\Theta_6$. Size-dependent breakage helped only together with shear aggregation. With Brownian aggregation, the fitted $\alpha$ went to zero, and $\Theta_4^{(\mathrm{B},\alpha)}$ reduced to $\Theta_4^{(\mathrm{B},0)}$. With shear aggregation but size-independent breakage, $\Theta_4^{(\mathrm{S},0)}$ could not be fitted, because the precipitates either grew without bound or the system became too stiff to solve.

A simple balance explains this interaction. Consider a precipitate of size $x$ in a suspension of solid volume fraction $\phi_\mathrm{s}$ and number concentration $N \approx \phi_\mathrm{s}/x^3$. Its collision rate per particle is $\beta N$, and its breakage rate is $S(x)$. For the Brownian kernel, $\beta$ is nearly size-independent, and the balance $\beta N = S$ gives
\begin{equation}
x \propto \phi_{\mathrm{s}}^{1/(3+3\alpha)}.
\label{eq:scaling brownian}
\end{equation}
For the shear kernel, $\beta \propto x^3$, so the collision rate $\beta N \propto G_\mathrm{eff}\phi_\mathrm{s}$ no longer depends on size. The balance with \eqref{eq:breakage rate} gives
\begin{equation}
x \propto \phi_{\mathrm{s}}^{1/(3\alpha)},
\label{eq:scaling shear}
\end{equation}
which has a finite solution only for $\alpha > 0$. Shear aggregation therefore needs size-dependent breakage to reach a stable precipitate size, as seen for $\Theta_4^{(\mathrm{S},0)}$. With Brownian aggregation, size-dependent breakage only weakens the concentration dependence of $x$, which the data did not support, and the fitted $\alpha$ went to zero.

Eqs.~\eqref{eq:scaling brownian} and \eqref{eq:scaling shear} also predict how the precipitate size scales with the feed concentration $c_0$, since $\phi_\mathrm{s} \propto c_0 - c_\mathrm{eq}$. This scaling was tested with the median size $d_{50}$ at 1500~s (Figure~\ref{fig:aggregation models}d). The slope of $\ln d_{50}$ versus $\ln c_0$ was 0.35 for the Brownian models, close to the value of $1/3$ from \eqref{eq:scaling brownian}. For $\Theta_4^{(\mathrm{S},\alpha)}$, it was 0.62, close to $1/(3\alpha) = 0.59$, while the data showed a slope of 1.01. $\Theta_4^{(\mathrm{S},\alpha)}$ thus predicted the direction and about two-thirds of the measured size reduction at 0.36~g/L. It also lowered the prediction error three-fold ($\mathcal{J}_\mathrm{pred} = 36.1$ versus 116.8). The predicted concentration at 0.36~g/L agreed with the data (Figure~\ref{fig:aggregation models}e), but the predicted PDF was broader and lower than measured (Figure~\ref{fig:aggregation models}fg). With a single training concentration, $\alpha$ was set by the PSD shape at 0.74~g/L, and the concentration scaling was an extrapolation. Experiments at more feed concentrations, which were beyond the scope of this study because of material constraints, would allow $\alpha$ to be identified more accurately. They would also pin the slope of $\ln d_{50}$ versus $\ln c_0$ more accurately. 

The best growth model failed this test more clearly (dash-dot lines in Figure~\ref{fig:aggregation models}). Fitted at 0.74~g/L, $\Theta_6$ predicted the 0.36~g/L batch with $\mathcal{J}_\mathrm{pred} = 380.0$, ten times the error of $\Theta_4^{(\mathrm{S},\alpha)}$ (Table~\ref{tab:model_selection}). Its high nucleation exponent ($u = 11.1$) made primary nucleation about 3000 times slower at half the supersaturation. The predicted concentration therefore stayed at the feed value for several tens of seconds, while the measured concentration had already dropped by 4~s (Figure~\ref{fig:aggregation models}e). With fewer nuclei sharing the solute, each precipitate grew larger. The predicted $d_{50}$ at 0.36~g/L was 132.4~\textmu{}m, larger than the 105.4~\textmu{}m predicted at 0.74~g/L, and the slope was $-0.32$ instead of the measured $+1.01$ (Figure~\ref{fig:aggregation models}d). The growth model thus predicted the wrong direction of the size change.

The effective shear rate was examined by profile likelihood: $G_\mathrm{eff}$ was fixed at values from $10^3$ to $10^9$~s$^{-1}$, and all other parameters were refitted at each value (Figure~S6). For COVID, the profile was flat from about $10^7$ to $10^9$~s$^{-1}$ and only bounded from below ($G_\mathrm{eff} \gtrsim 5\times10^6$~s$^{-1}$). The other parameters, the ratio $G_\mathrm{eff}/S_0$, and the prediction at 0.36~g/L did not change along this plateau, so $G_\mathrm{eff}$ was fixed at $10^7$~s$^{-1}$. For FLuc, the profile had an interior minimum, and $G_\mathrm{eff}$ was fitted. As $G_\mathrm{eff}$ is a lumped rate constant (Section~\ref{sec:kin aggregation}), only the ratio $G_\mathrm{eff}/S_0$ is interpreted below.

For FLuc, only the selected model and $\Theta_4^{(\mathrm{B},0)}$ were fitted. $\Theta_4^{(\mathrm{S},\alpha)}$ again had the lowest AIC ($-387.7$, compared with $-374.4$ for $\Theta_4^{(\mathrm{B},0)}$ and $-378.4$ for $\Theta_6$). $\Theta_4^{(\mathrm{S},\alpha)}$ was therefore selected and used for the rest of this paper.

\begin{table}[!htbp]
\centering
\caption{Model selection for COVID and FLuc. For each model, the table lists the number of fitted parameters $N_\mathrm{p}$, the objective $\mathcal{J}$ from \eqref{eq:objective} with $w = 0.2$, and the AIC from \eqref{eq:AIC}. For COVID, $\mathcal{J}_\mathrm{pred}$ is the objective of the batch at 0.36~g/L, predicted with the parameters fitted at 0.74~g/L. The effective shear rate of $\Theta_4^{(\mathrm{S},\alpha)}$ was fixed at $G_\mathrm{eff} = 10^7$~s$^{-1}$ for COVID, because its profile likelihood was flat, and was fitted for FLuc. Models marked NF were not fitted, n.c.\ marks values that were not computed, and -- marks a model that could not be fitted. The selected model is shown in bold.}
\label{tab:model_selection}
\small
\setlength{\tabcolsep}{4pt}
\begin{tabular}{llccccccc}
\toprule
 & & \multicolumn{3}{c}{Growth} & \multicolumn{4}{c}{Aggregation--breakage} \\
\cmidrule(lr){3-5}\cmidrule(lr){6-9}
 & & $\Theta_4$ & $\Theta_5$ & $\Theta_6$ & $\Theta_4^{(\mathrm{B},0)}$ & $\Theta_4^{(\mathrm{B},\alpha)}$ & $\Theta_4^{(\mathrm{S},0)}$ & $\boldsymbol{\Theta_4^{(\mathrm{S},\alpha)}}$ \\
\midrule
\multirow{4}{*}{COVID} & $N_\mathrm{p}$ & 4 & 5 & 6 & 4 & 5 & 4 & \textbf{4} \\
 & $\mathcal{J}$ & 0.972 & 0.612 & 0.485 & 0.508 & 0.508 & $\ge 2016$ & \textbf{0.447} \\
 & AIC & $-455.4$ & $-499.6$ & $-520.9$ & $-520.3$ & $-518.3$ & -- & $\mathbf{-533.1}$ \\
 & $\mathcal{J}_\mathrm{pred}$ (0.36~g/L) & n.c. & n.c. & 380.0 & 116.8 & 116.8 & -- & \textbf{36.1} \\
\midrule
\multirow{3}{*}{FLuc} & $N_\mathrm{p}$ & 4 & 5 & 6 & 4 & NF & NF & \textbf{5} \\
 & $\mathcal{J}$ & 12.54 & 7.18 & 2.02 & 2.18 & NF & NF & \textbf{1.87} \\
 & AIC & $-199.6$ & $-253.4$ & $-378.4$ & $-374.4$ & NF & NF & $\mathbf{-387.7}$ \\
\bottomrule
\end{tabular}
\end{table}

\subsubsection{Parametric Uncertainty Analysis} \label{sec: Uncertainty-analysis}

\begin{figure}[!htbp]
\centering
\includegraphics[width=\linewidth]{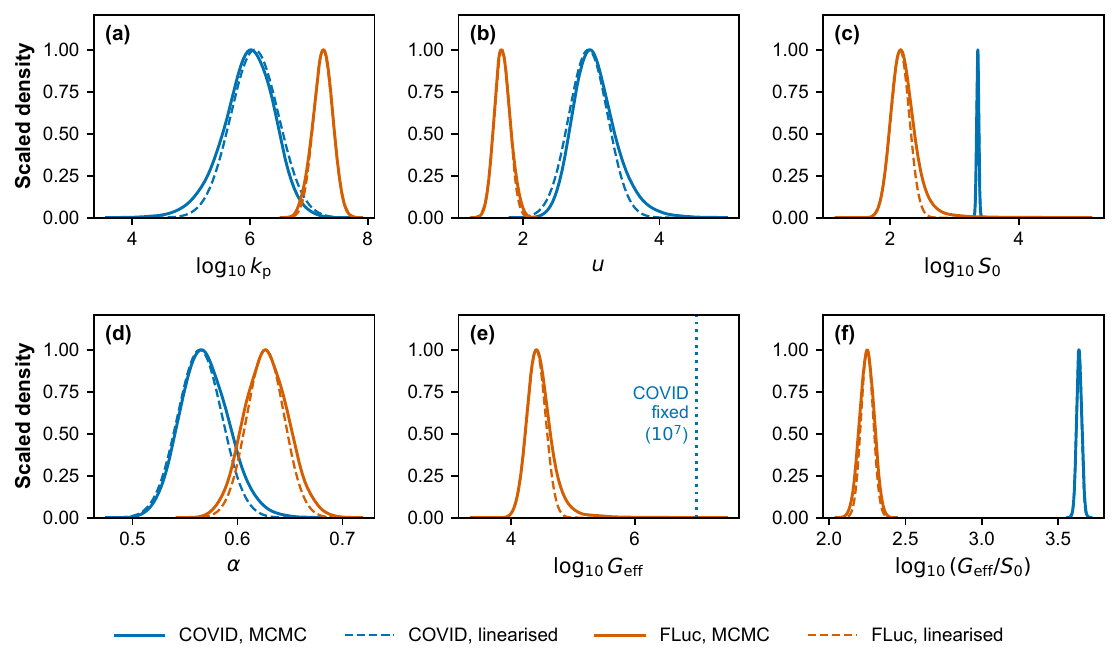}
\caption{Posterior distributions of the $\Theta_4^{(\mathrm{S},\alpha)}$ parameters for COVID (blue) and FLuc (orange): (a) $\log_{10}k_\mathrm{p}$, (b) $u$, (c) $\log_{10}S_0$, (d) $\alpha$, (e) $\log_{10}G_\mathrm{eff}$ (fixed at $10^7$~s$^{-1}$ for COVID, dotted line), and (f) $\log_{10}(G_\mathrm{eff}/S_0)$. Solid lines: MCMC (kernel density estimate); dashed lines: Gaussian from the linearized covariance at the best fit. Curves are scaled to a maximum of 1.}
\label{fig:Posterior Comparison}
\end{figure}

Figure~\ref{fig:Posterior Comparison} shows the MCMC posteriors of $\Theta_4^{(\mathrm{S},\alpha)}$, Table~\ref{tab:combined_uncertainty_intervals} lists the estimates, and Figures~S7 and S8 show the corner plots. All parameters had well-defined, single-peaked posteriors for both constructs. This was not the case for the growth model $\Theta_6$ (FLuc; Figure~S9). Its posterior of $u$ covered nearly the whole prior range (95\% interval 5.4--19.8), and $\log_{10}k_\mathrm{p}$ spanned about 30 decades along a ridge with $u$ ($r = -1.00$). The growth and secondary nucleation parameters were also strongly correlated with each other ($|r| = 0.87$--$0.98$). For $\Theta_4^{(\mathrm{S},\alpha)}$, the nucleation parameters $k_\mathrm{p}$ and $u$ were also strongly anti-correlated ($r = -0.96$ for COVID and $-0.99$ for FLuc). Despite this correlation, both were bounded: the 68\% intervals of $\log_{10}k_\mathrm{p}$ spanned less than one decade and those of $u$ less than 0.6 (Table~\ref{tab:combined_uncertainty_intervals}), whereas $u$ was nearly unbounded for $\Theta_6$. For FLuc, $S_0$ and $G_\mathrm{eff}$ were also correlated ($r = 0.99$). Their marginals had a long upper tail, and the posterior standard deviation was about twice the linearized standard error. Their ratio, however, was well determined, $G_\mathrm{eff}/S_0 = 177$ (68\% interval 159--196). For all other parameters, the posterior standard deviation was within 1.0--1.16 times the linearized standard error. The linearized covariance was therefore a fair summary, except for the upper tail of $S_0$ and $G_\mathrm{eff}$ for FLuc.

The breakage exponent was similar for both constructs ($\alpha = 0.57$ for COVID and 0.63 for FLuc), so that $S \propto x^{1.7}$--$x^{1.9}$. The constructs differed mainly in the ratio $G_\mathrm{eff}/S_0$, which was about 4300 for COVID and 180 for FLuc. The nucleation exponents were low ($u = 3.0$ for COVID and 1.7 for FLuc), lower than those of the growth model $\Theta_6$ ($u = 11.1$ for COVID, and $u = 5$ for FLuc; Table~\ref{tab:method_comparison}) and below those typical of crystallization (Section~\ref{sec: Priors from literature}). The high exponents for $b$ and $g$ of the growth model were thus likely a consequence of the missing aggregation. With such low exponents in the aggregation model, the fitted nucleation rate depended only weakly on supersaturation. $B_\mathrm{p} = k_\mathrm{p}s^u$ is therefore best read as an effective rate of molecular association that forms the smallest detectable precipitates, rather than as a classical nucleation rate.

\begin{table}[!htbp]
\centering
\caption{Best-fit parameters, MCMC posterior medians, and 68\% credible intervals ($x_{16\%}$, $x_{84\%}$) of $\Theta_4^{(\mathrm{S},\alpha)}$ for COVID and FLuc. For COVID, $G_\mathrm{eff}$ was fixed at $10^7$~s$^{-1}$. Prefactors are given as $\log_{10}$ ($k_\mathrm{p}$ in m$^{-3}$\,s$^{-1}$, $S_0$ and $G_\mathrm{eff}$ in s$^{-1}$).}
\label{tab:combined_uncertainty_intervals}
\small
\setlength{\tabcolsep}{4pt}
\begin{tabular}{llcccccc}
\toprule
 & & $\log_{10} k_\mathrm{p}$ & $u$ & $\log_{10} S_0$ & $\alpha$ & $\log_{10} G_\mathrm{eff}$ & $G_\mathrm{eff}/S_0$ \\
\midrule
\multirow{3}{*}{COVID} & Best fit & $6.08$ & $2.96$ & $3.362$ & $0.564$ & $7$ (fixed) & $4342$ \\
 & Median & $6.00$ & $3.03$ & $3.362$ & $0.568$ & -- & $4343$ \\
 & 68\% CI & $[5.55,\,6.39]$ & $[2.76,\,3.36]$ & $[3.345,\,3.378]$ & $[0.547,\,0.592]$ & -- & $[4185,\,4517]$ \\
\midrule
\multirow{3}{*}{FLuc} & Best fit & $7.24$ & $1.69$ & $2.16$ & $0.627$ & $4.40$ & $178$ \\
 & Median & $7.24$ & $1.69$ & $2.20$ & $0.627$ & $4.44$ & $177$ \\
 & 68\% CI & $[7.07,\,7.40]$ & $[1.59,\,1.81]$ & $[2.05,\,2.39]$ & $[0.607,\,0.648]$ & $[4.29,\,4.65]$ & $[159,\,196]$ \\
\bottomrule
\end{tabular}
\end{table}

\begin{figure}[!htbp]
\centering
\includegraphics[width=\textwidth]{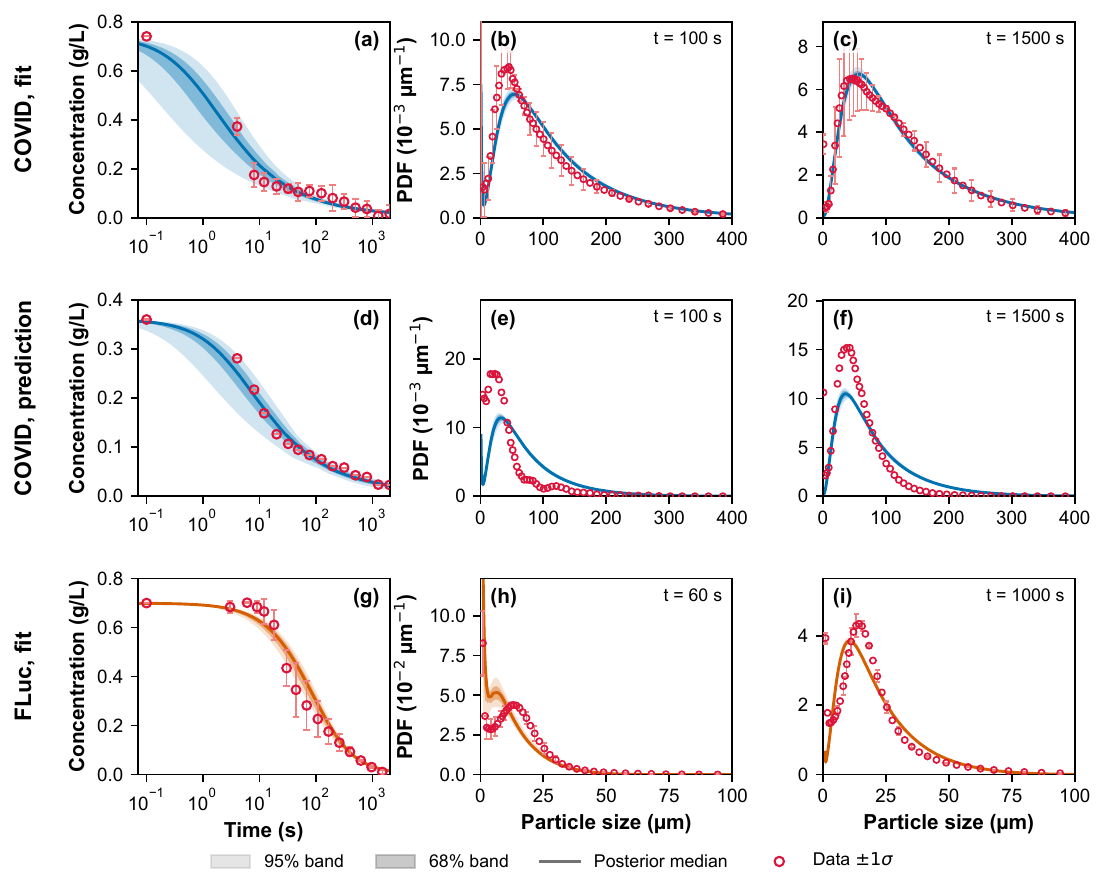}
\caption{Output uncertainty of $\Theta_4^{(\mathrm{S},\alpha)}$ propagated from the MCMC posterior, for the COVID fit at 0.74~g/L (top row), the COVID prediction at 0.36~g/L (middle row), and the FLuc fit (bottom row): (a, d, g) concentration versus time (logarithmic time axis), (b, e, h) PDF at an early time ($t = 100$~s for COVID, $t = 60$~s for FLuc), and (c, f, i) PDF at a late time ($t = 1500$~s for COVID, $t = 1000$~s for FLuc). Shaded regions are the 68\% and 95\% bands, and the solid line is the posterior median. Data are shown as mean $\pm1\sigma$.}
\label{fig:uncertainty COmbined}
\end{figure}

Figure~\ref{fig:uncertainty COmbined} shows the output uncertainty from 2000 posterior samples. The band was widest in the first 10--100~s, where nucleation set the rate of solute depletion, and it collapsed once the solute was depleted. For both fits (COVID at 0.74~g/L and FLuc), the bands were narrow and enclosed most data points. For the COVID prediction at 0.36~g/L, the band was also narrow. The prediction followed the measured concentration but did not reach the measured PDF peak (about 0.011 versus 0.015~\textmu{}m$^{-1}$ at 1500~s; Figure~\ref{fig:uncertainty COmbined}f). Because the band did not cover this gap, the gap was not caused by parameter uncertainty. It reflected what a single training concentration could identify. The breakage exponent $\alpha$ was set by the PSD shape at 0.74~g/L, so the change of size with feed concentration was extrapolated (Section~\ref{sec: aggregation models}).

\subsubsection{Comparison of mRNA Construct Kinetics}\label{sec :Comparison of kinetic rates}

\begin{figure}[h!]
\centering
\includegraphics[width=0.7\textwidth]{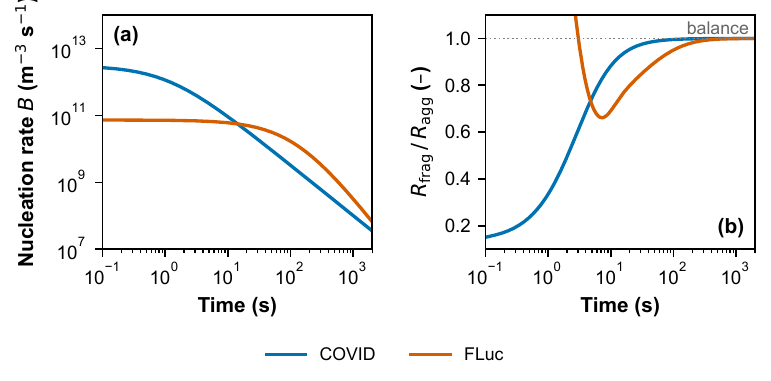}
\caption{Kinetics of $\Theta_4^{(\mathrm{S},\alpha)}$ for COVID (blue) and FLuc (orange): (a) nucleation rate $B_\mathrm{p} = k_\mathrm{p}s^u$ and (b) ratio of the breakage and aggregation event rates, $R_\mathrm{frag}/R_\mathrm{agg}$ (1 = balance), from \eqref{eq:event rates}. Rates are shown from the first concentration measurement (3~s).}
\label{fig:COVID vs FLuc}
\end{figure}

Figure~\ref{fig:COVID vs FLuc} compares the nucleation rate and the balance of aggregation and breakage for the two constructs. The aggregation and breakage event rates are
\begin{equation}
    R_\mathrm{agg}(t) = \frac{1}{2}\int_0^\infty\!\!\int_0^\infty \beta_\mathrm{S}(x,x')\,n(t,x)\,n(t,x')\,\mathrm{d}x\,\mathrm{d}x',
    \qquad
    R_\mathrm{frag}(t) = \int_0^\infty S(x)\,n(t,x)\,\mathrm{d}x,
    \label{eq:event rates}
\end{equation}
with $\beta_\mathrm{S}$ from \eqref{eq:kernels} and $S$ from \eqref{eq:breakage rate}. The factor 1/2 counts each colliding pair once. Each aggregation event removes one particle, and each binary breakage event adds one, so that $\mathrm{d}N/\mathrm{d}t = B_\mathrm{p} - R_\mathrm{agg} + R_\mathrm{frag}$, where $N = \int_0^\infty n\,\mathrm{d}x$ is the number of particles per unit volume (m$^{-3}$). All rates are numbers of events per m$^3$ per s.

Nucleation was faster at the start for COVID ($B_\mathrm{p} \approx 3\times10^{12}$ versus $7\times10^{10}$~m$^{-3}$\,s$^{-1}$ at 0.1~s) but lasted longer for FLuc, because of its lower exponent ($u = 1.7$ versus 3.0; Figure~\ref{fig:COVID vs FLuc}a). This explained the slower solute depletion of FLuc. For COVID, aggregation dominated the early dynamics, and breakage caught up until $R_\mathrm{frag} = R_\mathrm{agg}$ (Figure~\ref{fig:COVID vs FLuc}b). The ratio $R_\mathrm{frag}/R_\mathrm{agg}$ reached 0.9 at about 12~s and 0.99 at 60~s. For FLuc, the ratio first dropped to 0.66 at about 7~s, while nucleation was still strong, and approached the balance more slowly (0.91 at 60~s and 0.99 at 300~s). The precipitate size followed the balance of aggregation and breakage through $G_\mathrm{eff}/S_0$, which was about 4300 for COVID and 180 for FLuc. At a similar solid volume fraction and breakage exponent, \eqref{eq:scaling shear} gives $x \propto (G_\mathrm{eff}/S_0)^{1/(3\alpha)}$. A 24-fold higher ratio then corresponds to precipitates about five to six times larger, consistent with the measured difference in size (Section~\ref{sec: batch data}).

A possible explanation for the higher $G_\mathrm{eff}/S_0$ of COVID is chain length, as COVID mRNA has about twice as many nucleotides as FLuc mRNA. A longer chain can bridge between precipitates more easily, which would increase the collision efficiency and the precipitate strength. Differences in total charge and conformation \cite{dobrynin_scaling_1995} and in secondary structure, which is extensive in SARS-CoV-2 RNA \cite{rangan_rna_2020}, may also contribute. Comparing constructs of different length, charge, and secondary structure would test these explanations.

\subsection{Prediction of the Continuous Precipitator}
\label{sec:Cont}

The batch-fitted FLuc models were used to predict the outlet of the continuous tubular precipitator (Section~\ref{sec: DPFR methods}). The P\'eclet number measured for the precipitated phase on the same static-mixer precipitator from a recent study, was $\mathrm{Pe} \approx 500$ \citep{Bhawnani2026}. At this P\'eclet number, axial dispersion did not affect the outlet (SI Section~S4.3), so all models were evaluated with the plug-flow equations \eqref{eq:plug flow}.

\begin{figure}[!htbp]
\centering
\includegraphics[width=0.8\textwidth]{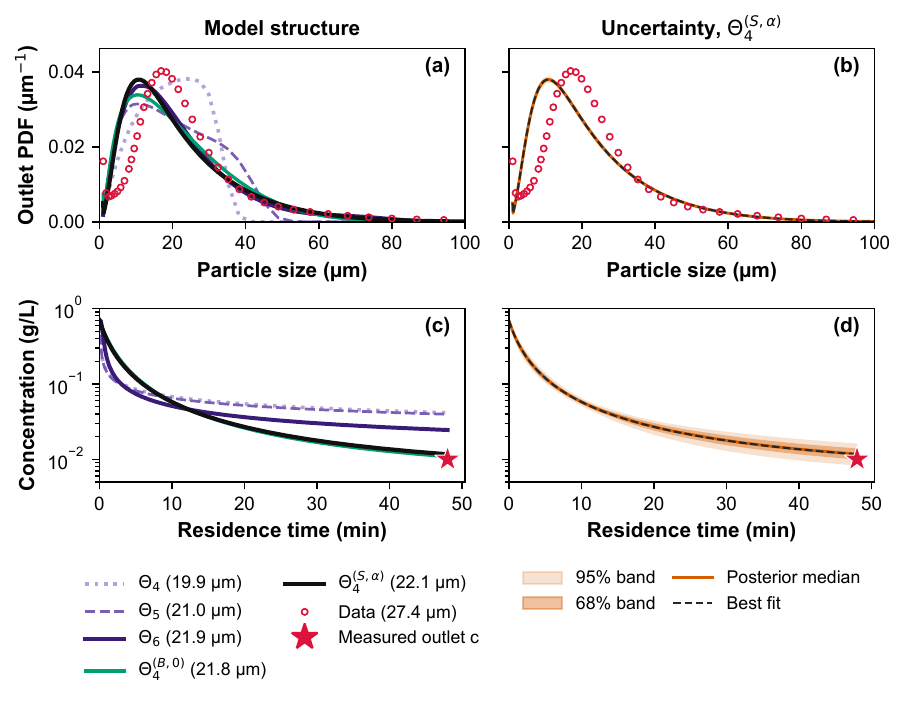}
\caption{Prediction of the continuous FLuc outlet from the batch-fitted models, in the plug-flow limit ($\tau = 48$~min). Top row: outlet PDF; bottom row: solute concentration along the reactor versus residence time, with the measured outlet concentration (star) at the residence time $\tau = 48$~min (sample collected 58~min after start-up). (a, c) Model structure: growth models $\Theta_4$--$\Theta_6$ and aggregation--breakage models $\Theta_4^{(\mathrm{B},0)}$ and $\Theta_4^{(\mathrm{S},\alpha)}$, with the predicted outlet mean length in the legend. (b, d) $\Theta_4^{(\mathrm{S},\alpha)}$ with the 68\% and 95\% bands from 1000 MCMC posterior samples; the dashed line is the best fit.}
\label{fig:continuous validation}
\end{figure}

The outlet concentration separated the two mechanisms (Figure~\ref{fig:continuous validation}c; Table~\ref{tab:cont pred}). The growth models predicted 0.025--0.041~g/L, 2.5 to 4 times the measured 0.01~g/L. Their high growth exponents ($g \approx 3$--$4$) made growth vanish as $s \to 0$, which slowed the final approach to equilibrium. The aggregation--breakage models predicted 0.011--0.012~g/L. For $\Theta_4^{(\mathrm{S},\alpha)}$, the 95\% band (0.008--0.016~g/L) contained the measured value (Figure~\ref{fig:continuous validation}d).

All models predicted similar outlet mean lengths of 19.9--22.1~\textmu{}m, below the measured 27.4~\textmu{}m. $\Theta_4^{(\mathrm{S},\alpha)}$ was the closest (22.1~\textmu{}m, an error of 19\%). Its 95\% band (21.6--22.7~\textmu{}m) was narrow and did not contain the measured value, so the size gap is structural rather than parametric. The predicted outlet PDF peaked at about 11~\textmu{}m, compared with about 17~\textmu{}m in the data (Figure~\ref{fig:continuous validation}ab). $\Theta_4$ and $\Theta_5$ gave distorted PDFs, with either a sharp cut-off or a pronounced shoulder at 30--40~\textmu{}m.

There are two possible explanations for the size gap. First, the hydrodynamics of the static mixer differ from those of the stirred vial, so the ratio $G_\mathrm{eff}/S_0$ in the tube may differ from its batch value. Second, the CLD was measured off-line after sampling, and aggregation may have continued in the collected sample.

This test showed that the batch-calibrated parameters transferred to continuous operation, with an accurate prediction of the outlet concentration and an outlet size within 20\% of the measured value.

\begin{table}[!htbp]
\centering
\caption{Predicted outlet mean length, absolute error, and outlet concentration for the continuous FLuc precipitator (plug-flow limit). The measured mean length was 27.40~\textmu{}m and the measured outlet concentration was 0.01~g/L. For $\Theta_4^{(\mathrm{S},\alpha)}$, the 95\% interval from the MCMC posterior is given in brackets.}
\label{tab:cont pred}
\begin{tabular}{cccc}
\toprule
\textbf{Model} & \textbf{Mean length (\textmu{}m)} & \textbf{Absolute error (\textmu{}m)} & \textbf{Outlet concentration (g/L)} \\
\midrule
$\Theta_4$ & 19.94 & 7.46 & 0.041 \\
$\Theta_5$ & 21.03 & 6.38 & 0.040 \\
$\Theta_6$ & 21.85 & 5.55 & 0.025 \\
$\Theta_4^{(\mathrm{B},0)}$ & 21.80 & 5.61 & 0.011 \\
$\Theta_4^{(\mathrm{S},\alpha)}$ & 22.14 [21.62, 22.75] & 5.26 & 0.012 [0.008, 0.016] \\
\midrule
Measured & 27.40 & -- & 0.01 \\
\bottomrule
\end{tabular}
\end{table}

\section{Conclusion}\label{sec: Conclusion}

This article develops PBMs for the continuous precipitation of mRNA for purification and calibrated them with in-line batch measurements of concentration and chord length for COVID and FLuc mRNA. Growth models with secondary nucleation fit the data, but only with unusually high supersaturation exponents, which point to a missing mechanism of aggregation and breakage. A model with primary nucleation, shear aggregation, and size-dependent breakage, $\Theta_4^{(\mathrm{S},\alpha)}$, fit both constructs best by the AIC. Shear aggregation and size-dependent breakage were only effective together: their balance set a stable precipitate size that grew with the solid volume fraction. All parameters of this model had well-defined posteriors. The model explained the roughly five-fold larger COVID precipitates by a 24-fold higher ratio of aggregation to breakage, $G_\mathrm{eff}/S_0$.

For the COVID batch at half the feed concentration, the model predicts the concentration profile accurately and captures about two-thirds of the measured reduction in precipitate size. Its error is three-fold lower than that of the Brownian aggregation model and ten-fold lower than that of the best growth model, which predicted an increase in precipitate size. For the continuous FLuc precipitator, the batch-calibrated model predicts the outlet concentration within its uncertainty band and the outlet mean chord length to within 20\%. These results show that batch experiments can be used to calibrate predictive models for continuous mRNA precipitation, which reduces the material needed during process development.

Remaining areas for improvement include a better description of the narrow precipitate size distributions measured at the lower feed concentration and at the continuous outlet. Experiments over a broader range of feed concentrations could allow a more direct estimate of the breakage exponent, and experiments at additional flow rates could show how the balance of aggregation and breakage depends on the hydrodynamics. With these refinements, the model and its uncertainty framework could support model-based design and control of continuous mRNA precipitation, including prediction of the outlet precipitate size distribution for downstream tangential flow filtration.

\section*{Acknowledgments}
This work was supported by the U.S. Food and Drug Administration under Contract 75F40122C00200. Krystian Ganko was additionally supported by the U.S. Department of Energy, Office of Science, Office of Advanced Scientific Computing Research, Department of Energy Computational Science Graduate Fellowship under Award Number DE-SC0022158.

\bibliography{Paper1/references_final_standardized}

\clearpage
\includepdf[pages=-]{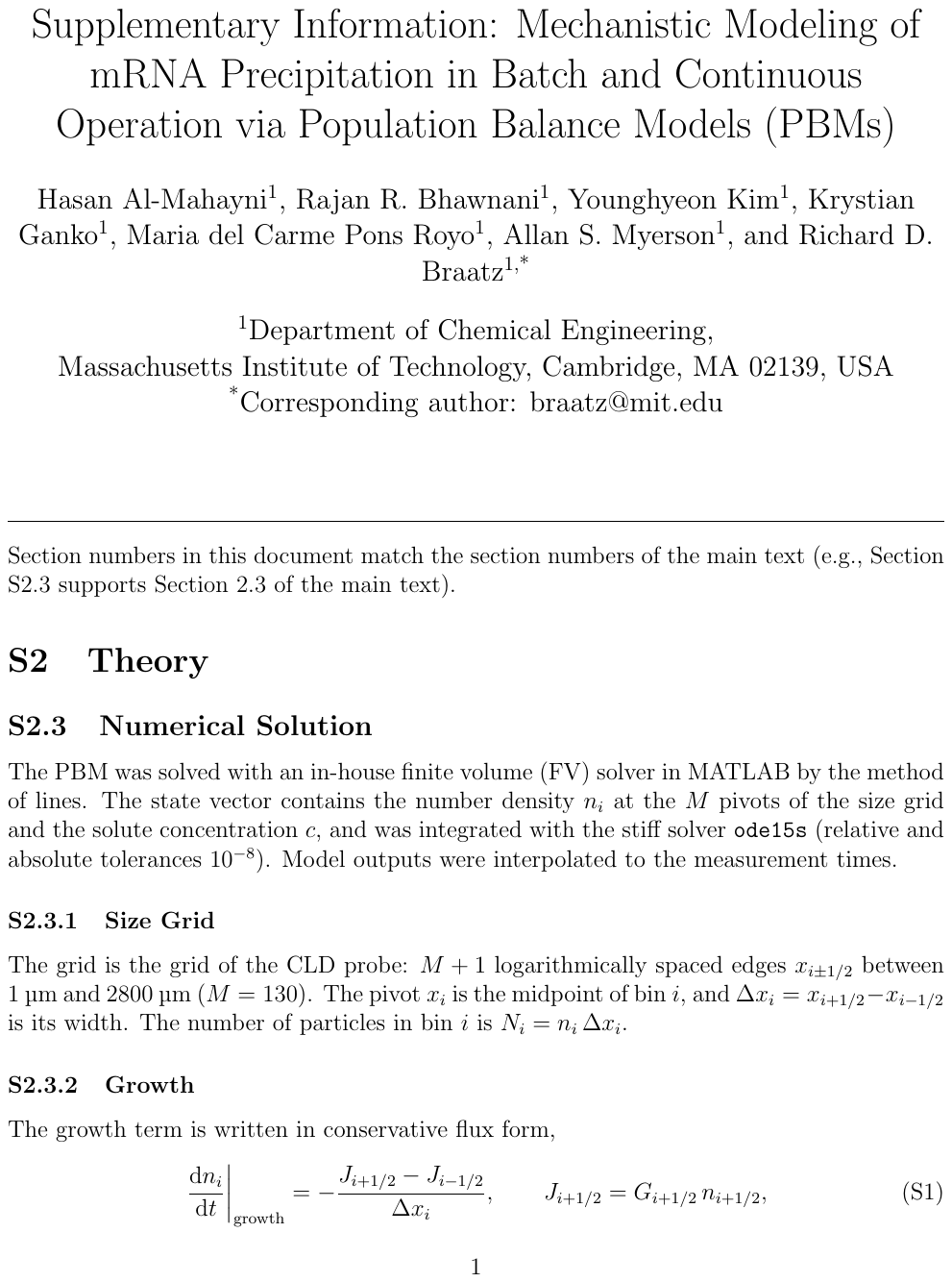}

\end{document}